\documentclass[dvips]{acta}
\usepackage{supertabular,lscape,epsfig}
\usepackage{amssymb}
\usepackage{amsmath}
\DeclareSymbolFont{ppa}{OT1}{ppl}{m}{it}
\DeclareMathSymbol{\vv}{\mathalpha}{ppa}{'166}

\newfont{\hb}{rphvb at 10pt}
\newfont{\hbo}{rphvbo at 10pt}
\newfont{\bitt}{rptmbi at 12pt}
\newfont{\bits}{rptmbi at 11pt}

\SetPages{313}{328}

\SetVol{58}{2008}

\begin{document}

\newcommand{\TabCapp}[2]{\begin{center}\parbox[t]{#1}{\centerline{
  \small {\spaceskip 2pt plus 1pt minus 1pt T a b l e}
  \refstepcounter{table}\thetable}
  \vskip2mm
  \centerline{\footnotesize #2}}
  \vskip3mm
\end{center}}

\newcommand{\TTabCap}[3]{\begin{center}\parbox[t]{#1}{\centerline{
  \small {\spaceskip 2pt plus 1pt minus 1pt T a b l e}
  \refstepcounter{table}\thetable}
  \vskip2mm
  \centerline{\footnotesize #2}
  \centerline{\footnotesize #3}}
  \vskip1mm
\end{center}}

\newcommand{\MakeTableSepp}[4]{\begin{table}[p]\TabCapp{#2}{#3}
  \begin{center} \TableFont \begin{tabular}{#1} #4 
  \end{tabular}\end{center}\end{table}}

\newcommand{\MakeTableee}[4]{\begin{table}[htb]\TabCapp{#2}{#3}
  \begin{center} \TableFont \begin{tabular}{#1} #4
  \end{tabular}\end{center}\end{table}}

\newcommand{\MakeTablee}[5]{\begin{table}[htb]\TTabCap{#2}{#3}{#4}
  \begin{center} \TableFont \begin{tabular}{#1} #5 
  \end{tabular}\end{center}\end{table}}

\newfont{\bb}{ptmbi8t at 12pt}
\newfont{\bbb}{cmbxti10}
\newfont{\bbbb}{cmbxti10 at 9pt}
\newcommand{\uprule}{\rule{0pt}{2.5ex}}
\newcommand{\douprule}{\rule[-2ex]{0pt}{4.5ex}}
\newcommand{\dorule}{\rule[-2ex]{0pt}{2ex}}
\begin{Titlepage}
\Title{Period Changes of LMC Cepheids\\ in the OGLE and 
MACHO Data\footnote{Based on observations obtained with the 1.3~m Warsaw
telescope at the Las Campanas Observatory of the Carnegie Institution of
Washington.}}
\Author{R.~~P~o~l~e~s~k~i}{Warsaw University Observatory, 
Al.~Ujazdowskie~4, 00-478~Warszawa, Poland\\
e-mail: rpoleski@astrouw.edu.pl}
\Received{October 18, 2008}
\end{Titlepage}
\vspace*{-12pt}
\Abstract{Pulsation period of Cepheids should change as stars evolve 
through the instability strip. Rates of these changes found by other
authors based on the decades-long $O-C$ diagrams show rather good agreement
with theoretical predictions. We have checked the variability on the scale
of a few years on the data recently published by the Optical Gravitational
Lensing Experiment (OGLE) for the Large Magellanic Cloud Cepheids and found
period changes for 18\% of fundamental mode and 41\% of first overtone
pulsators. It suggest the overtone pulsations are less stable than the
fundamental ones. For stars which had the cross-references in the MACHO
catalog we have checked if the period change rates derived from the OGLE
and the MACHO data are consistent. It was found that there is no correlation and
opposite signs of changes in both data sets are more common than the same
ones. Many $O-C$ diagrams show nonlinear period changes similarly as for
some stars the diagrams derived from the OGLE data only (spanning up to 4100
days) show random fluctuations. These fluctuations are common on the
long-term $O-C$ diagrams and we conclude they dominate the diagrams for the
timescales of a few thousand of days. The distributions of periods and
colors for all Cepheids and for those with statistically significant period
changes are the same. Times of maximum light obtained using the MACHO and
the OGLE data as well as the examples of $O-C$ diagrams are
presented.}{Stars: oscillations
-- Cepheids -- Magellanic Clouds}

\Section{Introduction}
Stars with masses higher than a few solar masses after main sequence phase
may cross the Cepheid instability strip in the Hertzsprung--Russell
diagram. Then, such a star becomes unstable to radial pulsations and can
be observed as a Cepheid. Detailed investigation on the Cepheid pulsation
properties can give constrains for stellar pulsation and evolution
theories. That is especially true for objects caught during the first
crossing of the instability strip \ie while being in the Hertzsprung gap.

Cepheids should change periods during each crossing of the instability
strip with the positive sign of the period change rate for odd and negative
for even crossings. Theoretically predicted timescales of period changes
($P/\dot{P}$) are between $10^4$ and $10^7$ years and should be measurable
on a few decades long $O-C$ diagrams (see \eg Zhou 1999 for a review). For
some stars there are times of maximum light observed for more than 100
years and detailed $O-C$ diagrams can be constructed. In this traditional
method data obtained by different observers using different telescopes,
filters and light detectors are used. The results of such investigations
were shown \eg by Berdnikov (1994), Berdnikov and Pastukhova (1994),
Berdnikov \etal (1997) and Berdnikov \etal (2004). They show generally good
agreement with evolutionary predictions (Turner \etal 2006).
\vskip9pt
The second method for measuring period changes is to compare periods found
in two distant epochs (\eg Pietrukowicz 2001, 2002, Bird \etal 2008). One
assumes the period change rate is constant or nearly constant and divides
the difference of periods by the time difference of two epochs obtaining
period change rate. If only one such value is estimated for each Cepheid
there is no way to verify if the assumption of constant period change rate
is true and all the results are uncertain. Deasy and Wayman (1985) used 3
epochs separated by $\approx44$ and $\approx10$ years. They found about
half of Cepheids to have variable rate of period changes.
\vskip9pt
One can use another method -- assume trial period $P$ and its change rate
$\dot{P}$, fold the data and evaluate some statistic to check if $P$ and
$\dot{P}$ were chosen properly. Extreme of this statistic should yield
correct $P$ and $\dot{P}$ values (\eg Kubiak \etal 2006, Pilecki \etal
2007). In this method one also assumes constant period change rate. The way
to check if it is true can be a comparison of the scatter in data folded
using estimated $P$ and $\dot{P}$ with photometric uncertainties. Data used
in that method should be transformed to the standard system, if they were
obtained with different telescopes or detectors.
\vskip9pt
In this work we rely on the Optical Gravitational Lensing Experiment (OGLE)
data. OGLE is a long-term wide-field microlensing survey. As a by-product
huge amount of homogeneous photometric data are obtained. Thus, the last
method described above can be used. Theoretically predicted Cepheids period
changes (Turner \etal 2006) should cause big enough phase shift since the
beginning of the OGLE-II in January 1997 for all first crossing Cepheids to
be detectable. It is also true for the Cepheids with long periods during
the second and the third crossing.
\vskip9pt
Soszyñski \etal (2008) published first part of the OGLE-III Catalog of
Variable Stars (OIII-CVS) which presents data for over 3000 single mode
Cepheids in the Large Magellanic Cloud (LMC). The accuracy of the period
change rate determination increases as a square of time base of
observations, so we decided to use also the MACHO data which cover years
1992--1999. Longer time-base give opportunity to check period change rate
variations.
\vskip9pt
In Section~2 we describe the photometric data and the methods used in
present analysis. The results are shown in Section~3 and their discussion
is given in Section 4. For convenience we refer O2 and O3 to the stars
for which the OGLE-II data do and do not exist. For all analyzed stars the
OGLE-III data are available.

\Section{Data and Methods}
The OGLE data have been collected with the 1.3-m Warsaw telescope located
at the Las Campanas Observatory. The camera is an eight chip SITe
$2048\times4096$ CCD mosaic. Each pixel is 15~$\mu$m and corresponding
scale is 0.26~arcsec/pixel. Full description of the setup was given by
Udalski (2003). The Difference Image Analysis (Alard and Lupton 1998, Alard
2000, Wo¼niak 2000, Udalski \etal 2008) was used to obtain the
photometry. Most of observations are taken through the $I$ filter and only
these data are used in this work. The OGLE-II and OGLE-III data for each
Cepheid were tied by Soszyñski \etal (2008). Before our analysis was
started we manually removed all the outlying points from the OIII-CVS
photometry.

The MACHO photometric data (Allsman and Axelrod 2001) were collected with
the 50-inch Great Melbourne Telescope located at the Mount Stromlo
Observatory. The dichroic element coupled to the telescope allowed
simultaneous observations in two nonstandard filters centered at 560 and
710~nm called $V_M$ and $R_M$, respectively. Each camera consisted of four
Loral $2096\times2096$ CCD chips. The scale was 0.62~arcsec/pixel. The PSF
photometry was obtained using So{\sc DoPHOT} software. Details were
published by Hart \etal (1996) and Alcock
\etal (1999). 

\hglue-4pt We included only these MACHO Cepheids for which there were 
cross-references given by Soszyñski \etal (2008). Significant number of the
MACHO data points were outlying. All points with error estimations higher
than 0.1~mag were removed. All points deviating from the average by more
than three times the dispersion were iteratively removed from the light
curves. Finally, Fourier series were fitted to the data and $3\sigma$
clipping procedure was applied. We checked by eye if the last procedure
have not removed points on the rising branch with steep slope. Several
times we decided to restore removed points. Stars with less than 100
photometric points were removed. The whole procedure was conducted
separately for the $V_M$ and $R_M$ data.

\MakeTable{lcrrrccrc}{9cm}{Number of stars in different data sets for 
fundamental~(F), first overtone~(1O) and both pulsation modes together}
{\hline
\noalign{\vskip3pt}
        & & & & & 
O2 & O2  & $R_M$ & O2, $R_M$ \\
& all & O2 & $R_M$ & $V_M$ & and & and & and & and\\
        & & & & & $R_M$ & $V_M$ & $V_M$ & $V_M$\\

\noalign{\vskip3pt}
\hline
\noalign{\vskip3pt}
F       & 1785 &  750 &  934 &  933 & 513 & 512 &  928 & 509 \\
1O      & 1161 &  510 &  575 &  571 & 327 & 323 &  570 & 323 \\
F or 1O & 2946 & 1260 & 1509 & 1504 & 840 & 835 & 1498 & 832 \\
\noalign{\vskip3pt}
\hline
\noalign{\vskip3pt}
\multicolumn{9}{p{9cm}}{All
objects are present in OGLE-III data. $R_M$ and $V_M$ stands for the MACHO
$R_M$ and $V_M$ filter data.}
}

We have taken the single mode fundamental (F) and first overtone (1O)
classical Cepheids from the OIII-CVS. Objects with additional
periodicities, mean brightness variability as well as ones with poor
phase-coverage near maximum light (mainly due to small number of data or
period close to integer number of days) were removed. The period given
there ($P_C$) for each object was used as starting value for our
calculations. Table~1 summarizes number of stars in different data sets.
We had on average 754, 384, 843 and 988 photometric points spanning 3920,
2370, 2680 and 2700 days for O2, O3, MACHO $R_M$ and MACHO $V_M$ data,
respectively.

\subsection{$O-C$ Method}
The data obtained by the sky surveys such as OGLE are hard to use in
traditional $O-C$ analysis. Typical cadence is one measurement per night
and only for the longest period Cepheids more than 10 points are obtained
during one cycle and traditional method of maximum time determination can
be applied. One can fold points from longer period of time and find $O-C$
value using Hertzsprung method (Berdnikov 1992) instead, \ie by shifting
the phase of mean light curve to minimize the sum of squared differences
between the observed brightness and the mean light curve. The mean light
curve was constructed by fitting Fourier series with the number of terms
set between 4 and 12 and chosen for each star separately to minimize
$\chi^2$ per degree of freedom. The terms with coefficients smaller than 3
times their errors were neglected.

If a star changes its period rapidly then several years long data can show
high scatter when folded with the best constant period. To avoid this
problem we first folded the light curve with $P_C$ and $\dot{P}=0$ and
calculated one $O-C$ value for each observing season. If there was evidence
for high period change rate on the resulting $O-C$ diagram we calculated
$P_0$ and $\dot{P}_0$ determined from each data set separately from the
parabolic fit. These values were used to obtain the new light curve and the
new Fourier fit.

In the final stage we divided the data from each observing season into
chunks. The number of data points in each observing season was divided by
10 for OGLE and 20 for MACHO data. Number of chunks for a given observing
season was equal to the integral part of the result. Each chunk contained
consecutive points which number was from 10 to 19 for the OGLE data and
from 20 to 39 for the MACHO data. Each chunk was used to find one $O-C$
value. The significance of parabolic term in the least squares fit to the
resulting $O-C$ diagram was checked using the F-test (Pringle 1975) with
1\% significance level and from the parabolic fit the values $P_1$ and
$\dot{P}_1$ were found. For stars with high amplitude variations not only
the phase was shifted but also the relative amplitude was fitted for each
chunk.

\subsection{Fourier Fitting}
In the second method used to search for period changes data were folded
according to the trial period $P$ and its change rate $\dot{P}$. Then,
Fourier series was fitted to the folded light curve and $\chi^2$ was
calculated. The trial values $P_2$ and $\dot{P}_2$ which minimize $\chi^2$
were found using the simplex method (Press \etal 1992). The Fourier fitting
was performed in the same way as in the $O-C$ method. The procedure of
\begin{figure}[htb]
\centerline{\includegraphics[width=12cm]{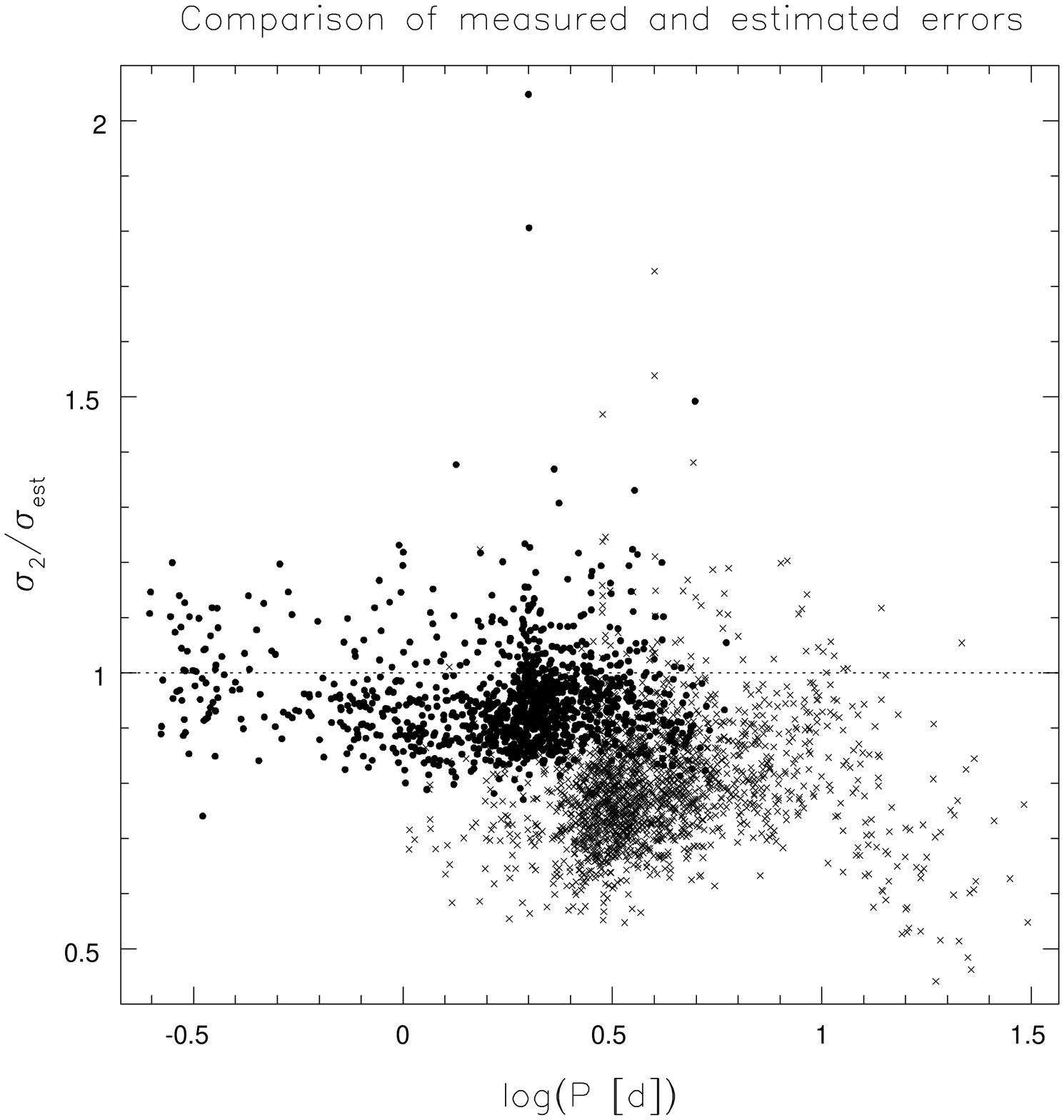}}
\FigCap{The ratio of period change rate error found on the OGLE data using 
Monte Carlo simulations and the Fourier fitting and estimated theoretically
for 1784 fundamental mode (crosses) and 1161 first overtone (dots) Cepheids
as a function of period. Cepheid with period 2.9990886~d and
$\sigma_2/\sigma_{\rm est}=5.3$ was omitted.}
\end{figure}
fitting Fourier series is linear and thus it is much easier to
perform. Only two parameters ($P$ and $\dot{P}$) were fitted by the simplex
method. If the $O-C$ method showed significant period change for a given
star then values $P_1$ and $\dot{P}_1$ were also used as starting points
for simplex. Errors in $\dot{P}$ were evaluated by 1000 iterations of Monte
Carlo simulations. We empirically found they do not depend on $\dot{P}$
assumed for the simulation. Henceforth, 3$\sigma$ criterion was applied to
check significance of $\dot{P}_2$.
Pilecki \etal (2007) proposed that estimation of the minimum error of
period change rate ($\sigma_{\rm est}$) for a star with the sinusoidal
light curve observed $N$ times uniformly distributed over time $T$ can be
given by
$$\sigma_{\rm est}\approx \frac{12P^2_{\rm sin}}{N^{1/2}T^2}
\cdot\frac{\sigma}{A}\eqno(1)$$
where $\sigma$ is photometric error, $P_{\rm sin}$ -- period of a sinusoid
(not to be confused with orbital period of a binary used by Pilecki \etal
2007) and $A$ -- full-amplitude.  Fig.~1 shows the ratio of this estimation
and values found by the Fourier fitting for the OGLE data ($\sigma_2$) as a
function of period separately for F and 1O Cepheids. Values of $A$ and
$\sigma$ were found using the light curves folded with $P_2$ and
$\dot{P_2}$. The points with high values correspond to stars with periods
close to integer number of days or those with large gaps in data. The
difference between F and 1O pulsators and change of this ratio with period
are caused by changes in the light curve shape (\cf Fig.~1 in Soszyñski
\etal 2008). Brightness of the longest period Cepheids changes very rapidly
on rising branch and a small period change causes big changes between
brightness predicted with assumption of constant period and true one. Thus,
some objects may have $\sigma_2$ much smaller than $\sigma_{\rm est}$. The
same figure for MACHO data is very similar, although they have smaller
seasonal gaps and the data are distributed more uniformly. We conclude that
the above estimation can be used to check what accuracy of the period
change rate can be achieved with given data.

\subsection{Comparison of Methods}
\begin{figure}[htb]
\includegraphics[width=12cm]{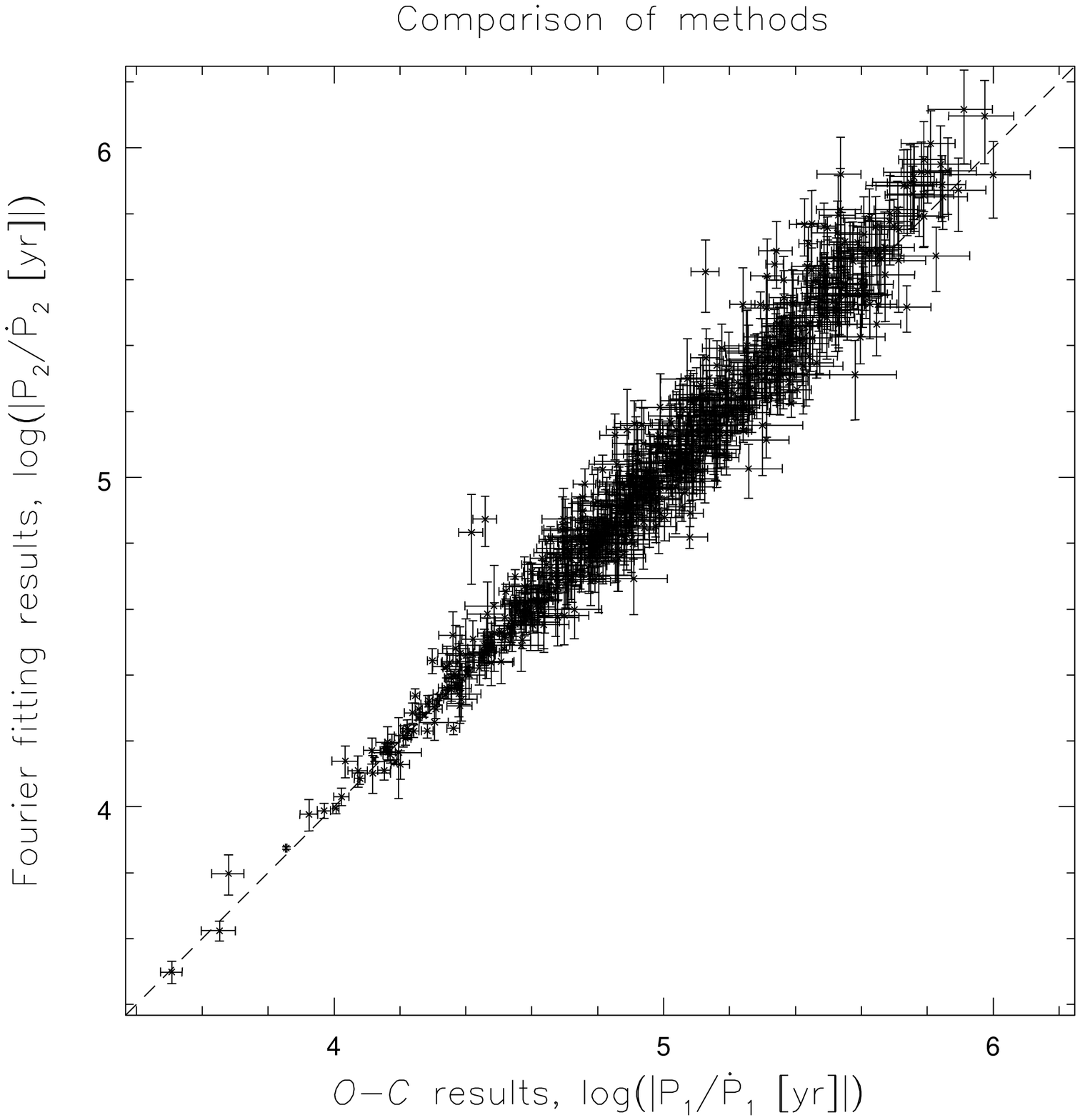}
\FigCap{Comparison of period derivatives found on OGLE data for 655
Cepheids using two different methods.}
\end{figure}
For the OGLE data exactly the same number of changing period Cepheids was
found using each of the methods (803 \ie 27\% of the sample). However, out
of them only 655 objects where found using both methods which is 22\% of
the sample and 69\% of Cepheids with period changes found using either
method. The only case for which the period change rates have opposite signs
was OGLE-LMC-CEP-2176, which is close to the saturation limit of the OGLE
photometry. The results from two methods for another 33 out of 655 Cepheids
were inconsistent in the meaning of the $3\sigma$ limit, in most cases due
to some additional variability on the $O-C$ diagrams. 

Evolutionary models predict that the period change rates for the first
crossing Cepheids are at least two times larger at given period than for
the third crossing Cepheids (Turner \etal 2006) with $\log{P}<1.1$ (98\% of
our sample). Fig.~2 compares the timescales of period changes derived from
the OGLE data using both methods described in the Sections 2.1 and
2.2. Comparison of the MACHO $R_M$ and $V_M$ results from the Fourier
fitting method shows the same signs and very similar rates for all
objects. These findings ensure that the methods are sufficient to
distinguish between evolutionary changes caused by different crossings of
the instability strip.

\Section{Results}
\begin{figure}[p]
\includegraphics[width=12.5cm]{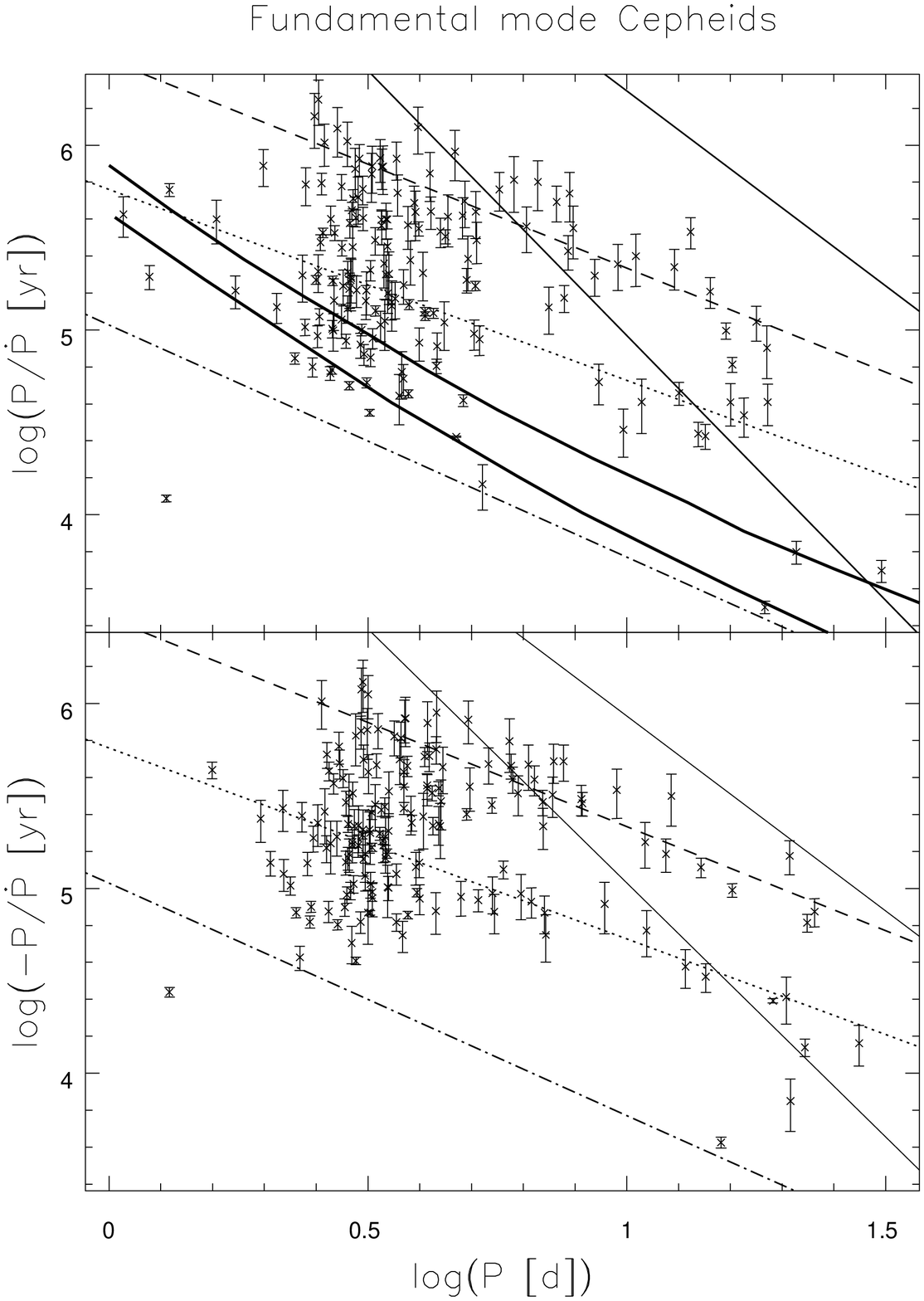}
\FigCap{Timescales of period change rates for fundamental mode Cepheids 
derived from OGLE data using Fourier fitting. {\it Upper} and {\it lower
panels} show positive and negative period changes, respectively. Solid
lines mark regions within which theoretically predicted (\cf Turner \etal
2006) evolutionary period changes for different crossings of the
instability strip should be found: thick lines mark the first crossing
region, thin lines mark the second and the third crossing region. Lines end
in the same points as in Fig.~2 in Turner \etal (2006). Dotted lines show
linear fit to detection limits for O2 (observed by OGLE-II and OGLE-III)
and dash lines for O3 (observed by OGLE-III and not observed by OGLE-II)
stars. Dash-dotted lines show thermal timescale.}
\end{figure}
\begin{figure}[htb]
\includegraphics[width=12.5cm]{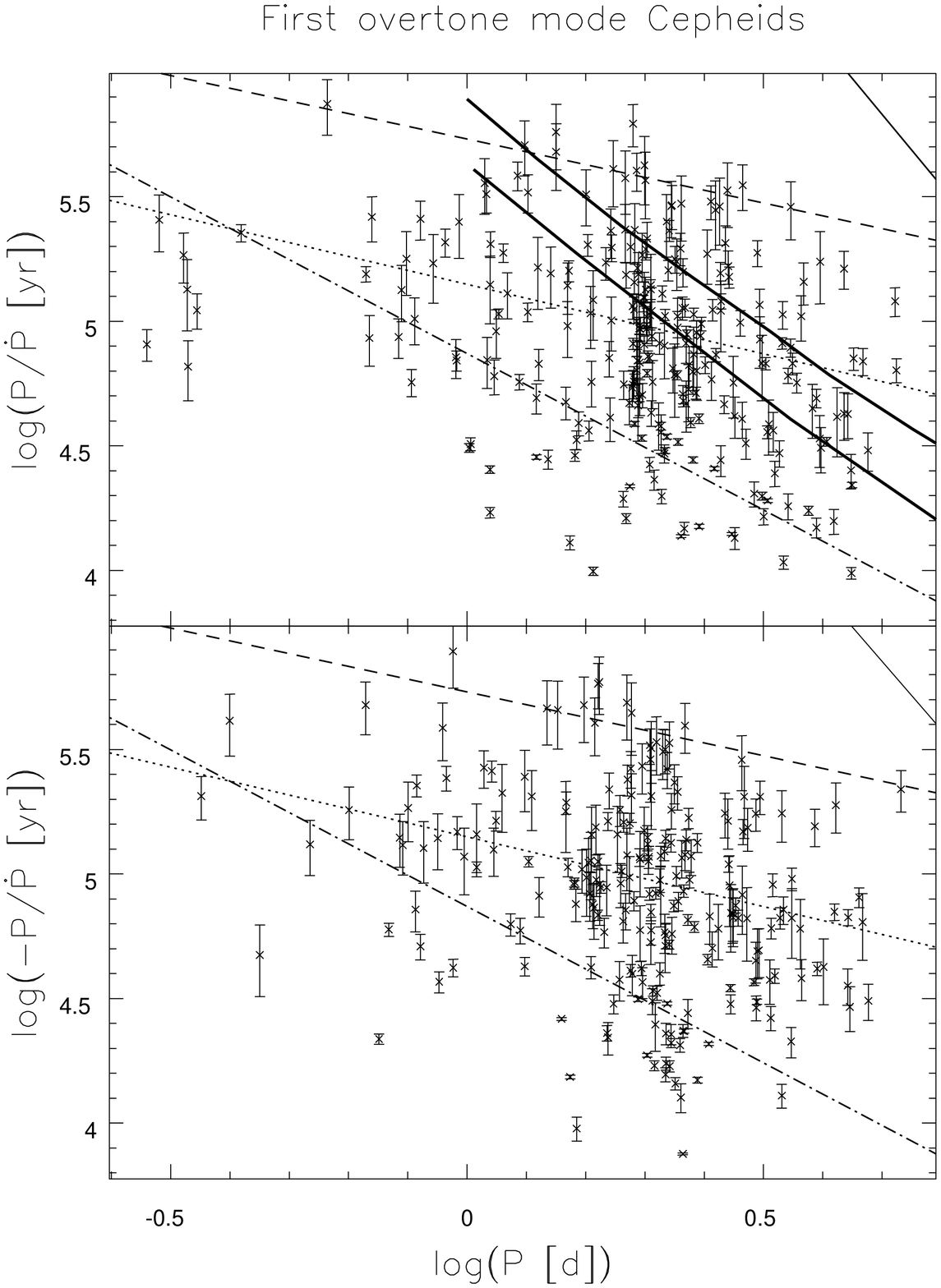}
\FigCap{Similar to Fig.~3 for 1O pulsators. Theoretical predictions 
are the same. For the second and the third crossing regions only the lower
limits are shown. Thermal timescale was shifted by $-0.125$ in $\log(P)$.}
\end{figure}
Fourier fitting analysis of the OGLE data showed period changes for 18\%
fundamental mode and for 41\% first overtone Cepheids. Logarithm of the
timescales of period changes ($\log\left|P_2/\dot{P}_2\right|$) as a
function of the logarithm of period is shown in Figs.~3 and 4 for F and 1O
pulsators, respectively. Solid lines mark theoretical predictions for the
first, second and third crossing of the instability strip taken from Turner
\etal (2006). Dash-dotted lines show thermal timescales (equal to $GM^2/RL$
where $G$ is gravitational constant, $M$ -- mass of the star, $R$ -- its
radius and $L$ -- luminosity) evaluated on the basis of Alibert \etal
(1999) models for the LMC metallicity. The number of objects falling in the
first crossing regions is striking. The time spent in the Hertzsprung gap
is so short that one per around a hundred Cepheids should be caught during
that phase of stellar evolution. We do not know any confirmed first
crossing Cepheid, last candidates being Polaris and DX~Gem (Turner \etal
2006), two LMC triple mode Cepheids (Moskalik and Dziembowski 2005),
shortest period Cepheids ($P<1.7$~d for LMC; \cf Alibert \etal 1999
Table~7) and almost all 1O/2O Cepheids (Baraffe \etal 1999). Moreover,
there are stars with the same timescale of period changes as for the first
crossing, but with negative $\dot{P}$ what is in conflict with the
evolutionary predictions. In fact, in both of these figures the
distribution of points is the same for positive and negative period
changes. Also the period and ($V-I$) color (Fig.~5) distributions of F and
1O Cepheids are the same for whole sample and stars with significant period
changes.

\begin{figure}[htb]
\centerline{\includegraphics[width=10cm]{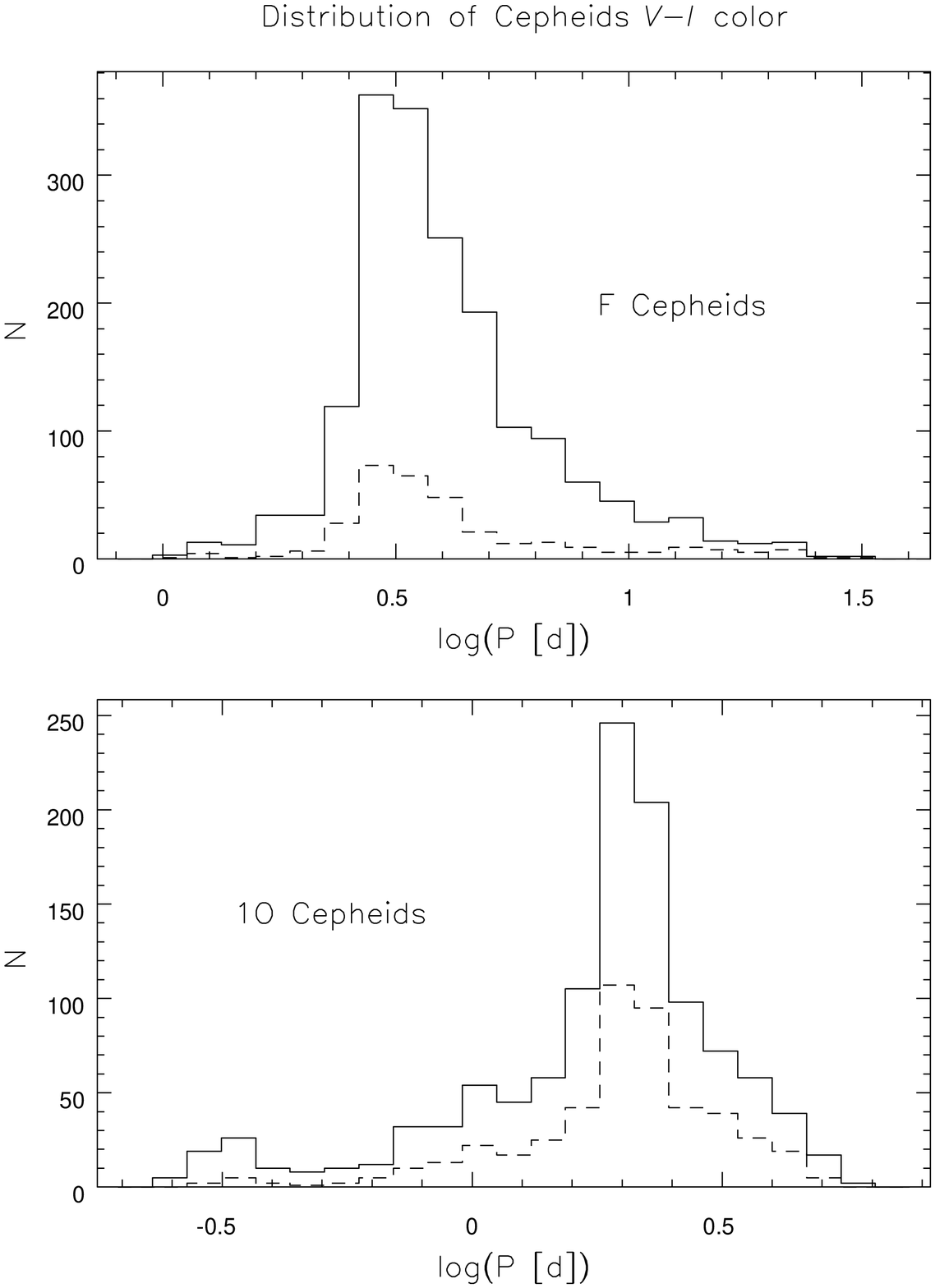}}
\FigCap{Distributions of $V-I$ color for F ({\it top}) and 1O ({\it bottom 
panel}) Cepheids. Distributions for all Cepheids in a given mode are shown
by solid lines and for changing period ones with dashed lines.}
\end{figure}

\MakeTable{llrrr}{12cm}{Percentage of all ($p$), positive ($p_+$) and negative 
($p_-$) period changes in different data sets}
{\hline
\noalign{\vskip3pt}
Data set & Puls. mode & \multicolumn{1}{c}{$p$} & 
\multicolumn{1}{c}{$p_+$} & 
\multicolumn{1}{c}{$p_-$} \\
\noalign{\vskip3pt}
\hline
\noalign{\vskip3pt}
OGLE    & 	   F & $18.1\pm0.9$\% & $ 8.9\pm0.7$\% & $ 9.2\pm0.7$\% \\
OGLE    & 	  1O & $41.3\pm1.4$\% & $21.9\pm1.2$\% & $19.5\pm1.2$\% \\
MACHO $R_M$ & 	   F & $11.1\pm1.0$\% & $ 4.0\pm0.6$\% & $ 7.2\pm0.8$\% \\
MACHO $R_M$ & 	  1O & $36.9\pm2.0$\% & $16.2\pm1.5$\% & $20.7\pm1.7$\% \\
MACHO $V_M$ & 	   F & $15.0\pm1.2$\% & $ 5.6\pm0.8$\% & $ 9.4\pm1.0$\% \\
MACHO $V_M$ & 	  1O & $43.8\pm2.1$\% & $19.8\pm1.7$\% & $24.0\pm1.8$\% \\
\noalign{\vskip3pt}
\hline
\noalign{\vskip3pt}
average &	   F & $15.0\pm0.7$\% & $ 6.0\pm0.4$\% & $ 8.6\pm0.5$\% \\
average &	  1O & $40.8\pm1.1$\% & $19.7\pm0.9$\% & $20.8\pm0.9$\% \\
\noalign{\vskip3pt}
\hline
}
Table~2 shows the incidence rates of period changing Cepheids. Errors were
calculated assuming Poisson distribution of the number of Cepheids with
period changes. In all cases first overtone pulsators show period changes
more often than the fundamental mode ones, with the average ratio of
incidence rates equal to 2.7. Timescales for 1O also group closer to
thermal timescale than for F Cepheids (Figs.~3 and 4). It may be
interesting to compare these values with the incidents rate of Blazhko
Cepheids: 4\% for F and 28\% for 1O (Soszyñski \etal 2008). Both results
suggest higher instability of the first overtone oscillations. For all data
sets and modes, except for the OGLE data for first overtone pulsators, the
negative period changes are more common than the positive ones. We found
the observational errors in $\dot{P}$ to be $\approx1.3$ times larger for
1O than for F Cepheids what is caused mainly by smaller amplitude of 1O. As
a consequence the observational limit in $\log\left(|P/\dot{P}|\right)$ is
smaller for 1O. For stars with significant period changes the average
modulus of $\dot{P}/\sigma_{\dot{P}}$ is equal to $7.3\pm0.5$ for F and
$11.1\pm0.6$ for 1O pulsators.

The period changes have the same signs for 23 of F (53 of 1O) and opposite
for 36 (87) Cepheids when comparing OGLE and MACHO $R_M$ Fourier fitting
results. OGLE and MACHO $V_M$ give 24 (65) the same and 48~(100) opposite
results, respectively.

\MakeTable{ccccc}{12.5cm}{Sample part of {\sf list.dat} file}
{\hline
\noalign{\vskip3pt}
Cpeheid ID & $P$ [d] & $\sigma_P$ [d] & Puls. mode & Remarks \\
\noalign{\vskip3pt}
\hline
\noalign{\vskip3pt}
OGLE-LMC-CEP-0771 & 2.1516256 & 0.0000017 & 1O & V,R,O \\
OGLE-LMC-CEP-0772 & 3.0735245 & 0.0000019 &  F & V,R,O \\
OGLE-LMC-CEP-0773 & 4.1847753 & 0.0000052 &  F & V,R   \\
OGLE-LMC-CEP-0774 & 3.1233001 & 0.0000177 & 1O &       \\
OGLE-LMC-CEP-0775 & 1.8634008 & 0.0000061 & 1O &       \\
OGLE-LMC-CEP-0776 & 6.8937966 & 0.0000062 &  F & O     \\
OGLE-LMC-CEP-0777 & 2.9979476 & 0.0000043 &  F &       \\
OGLE-LMC-CEP-0778 & 3.0904206 & 0.0000013 &  F & V,R,O \\
OGLE-LMC-CEP-0779 & 2.5607125 & 0.0000014 &  F & O     \\
OGLE-LMC-CEP-0780 & 3.8269168 & 0.0000035 &  F & V,R   \\
\noalign{\vskip3pt}
\hline
}
Using previously determined $O-C$ values we calculated times of maximum
light for each star. These data are available from the Internet archive:
\begin{center}
{\it ftp://ftp.astrouw.edu.pl/ogle/ogle3/OIII-CVS/lmc/cep/max\_time/}
\end{center}
The list of analyzed Cepheids is given in file {\sf list.dat} and contains
OIII-CVS ID, mean period and its error, pulsation mode and remarks which
may be: $V$, $R$, $O$, if MACHO $V_M$, MACHO $R_M$ and OGLE-II maxima were
calculated, and $A$ if the amplitude was variable in at least one of the
data sets. Mean periods were determined for each of the data set separately
using software {\sc Tatry} (Schwarzenberg-Czerny 1996). The MACHO
periods were then averaged and resulting values were averaged with the OGLE
periods to obtain mean periods. The application of other periods for
construction of the $O-C$ diagram is also possible, but would cause very
high or very low $O-C$ values making visual inspection harder and in
extreme cases leading to wrong interpretation.

\MakeTable{ccc}{12.5cm}{Sample part of {\sf OGLE-LMC-CEP-0772.max} file}
{\hline
\noalign{\vskip3pt}
HJD of maximum light & Error & Source \\
\noalign{\vskip3pt}
\hline
\noalign{\vskip3pt}
2450835.3347 & 0.0065 & O \\
2450847.6335 & 0.0081 & O \\
2450866.0462 & 0.0056 & V \\
2450872.2166 & 0.0063 & O \\
2450875.2750 & 0.0102 & R \\
2450912.1596 & 0.0185 & O \\
2451053.5473 & 0.0123 & V \\
2451078.1476 & 0.0069 & O \\
2451096.5588 & 0.0136 & R \\
2451111.9531 & 0.0050 & O \\
\noalign{\vskip3pt}
\hline
}

Files {\sf OGLE-LMC-CEP-NNNN.max}, where NNNN stands for the OIII-CVS
catalog number, contain in consecutive columns: time of maximum light
(HJD), its error (days) and data source identification ($O$, $R$ and $V$
for OGLE, MACHO $R_M$ and MACHO $V_M$, respectively). Tables~3 and
4 present sample parts of {\sf list.dat} and one of the {\sf
OGLE-LMC-CEP-NNNN.max} files. Calculated differences of times of maximum
light are reliable for each data source but may be subject to the shifts in
zero points of up to 0.1 day. They may be caused by different times of
maximum light in different filters or imperfections in fitting and finding
maximum of the Fourier series. Similar differences are found in the
traditional $O-C$ diagrams (\eg Berdnikov and Pastukhova 1994 and their
Table~3). Also analysis of the data for 671 Cepheids for which we have
MACHO times of maximum light, but we do not have the OGLE-II ones, has to
be done with caution in view of a possible epoch counting error. Presented
data may be added to the existing $O-C$ diagrams and/or tested for presence
of random fluctuations using methods described \eg by Lombard (1998) or
Berdnikov \etal (2004).

\begin{figure}[htb]
\includegraphics[width=11cm]{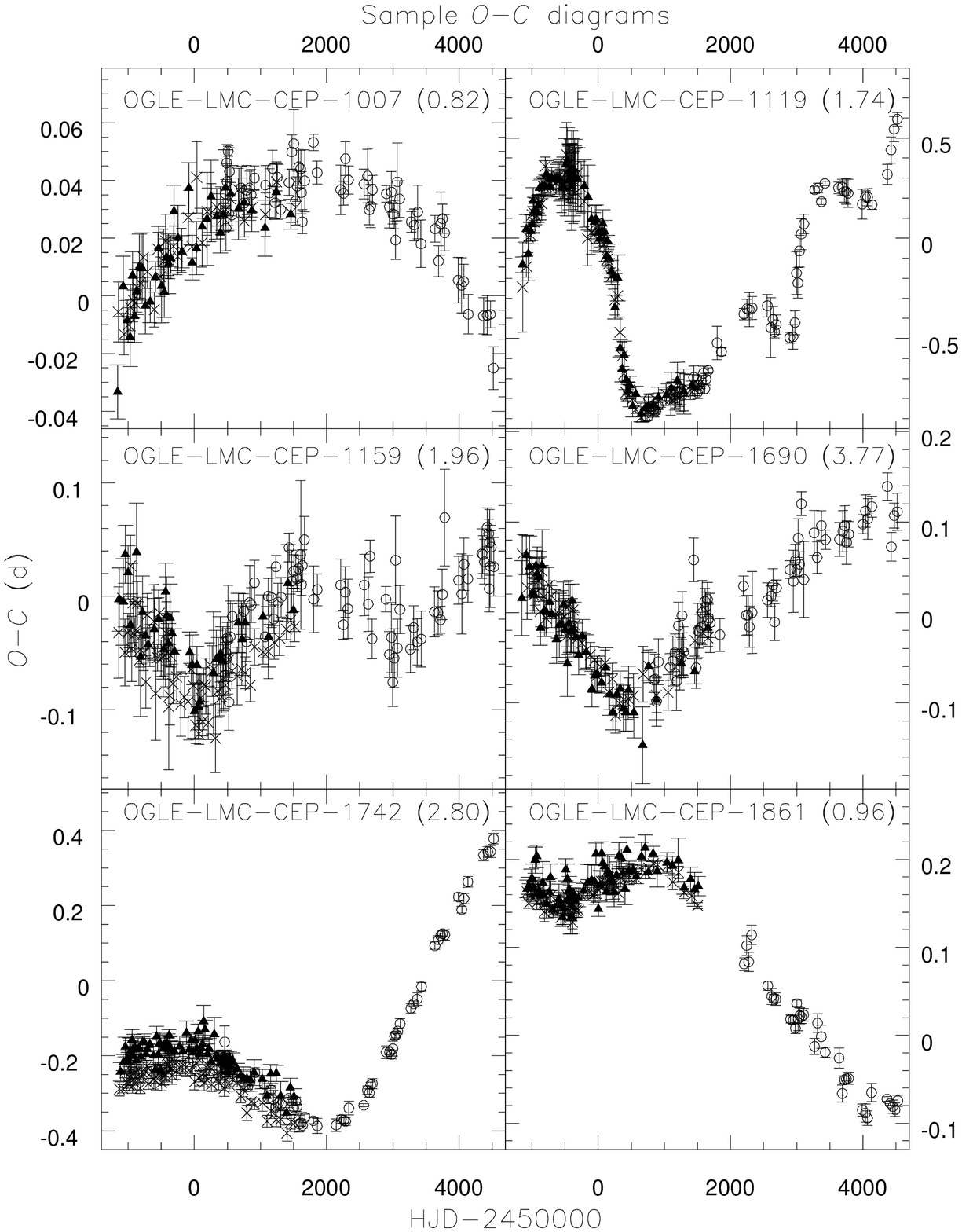}
\FigCap{Exemplary $O-C$ diagrams for the LMC Cepheids. Filled triangles, 
crosses and open circles correspond to MACHO $R_M$, MACHO $V_M$ and OGLE
data, respectively. OIII-CVS ID and rounded period (days) are given above
{\it each panel}. The abscissa is common for {\it all panels}, the ordinate
is given on left or right of {\it each panel}. OGLE-II and OGLE-III
data start at 460 or 740 and 2200, respectively. MACHO data end at 1500. No
filter dependent shifts for the $O-C$ values were
applied. OGLE-LMC-CEP-1742 and OGLE-LMC-CEP-1861 are examples of Cepheids
changing period in opposite directions in the OGLE and MACHO data.}
\end{figure}
\begin{figure}[htb]
\includegraphics[width=11cm]{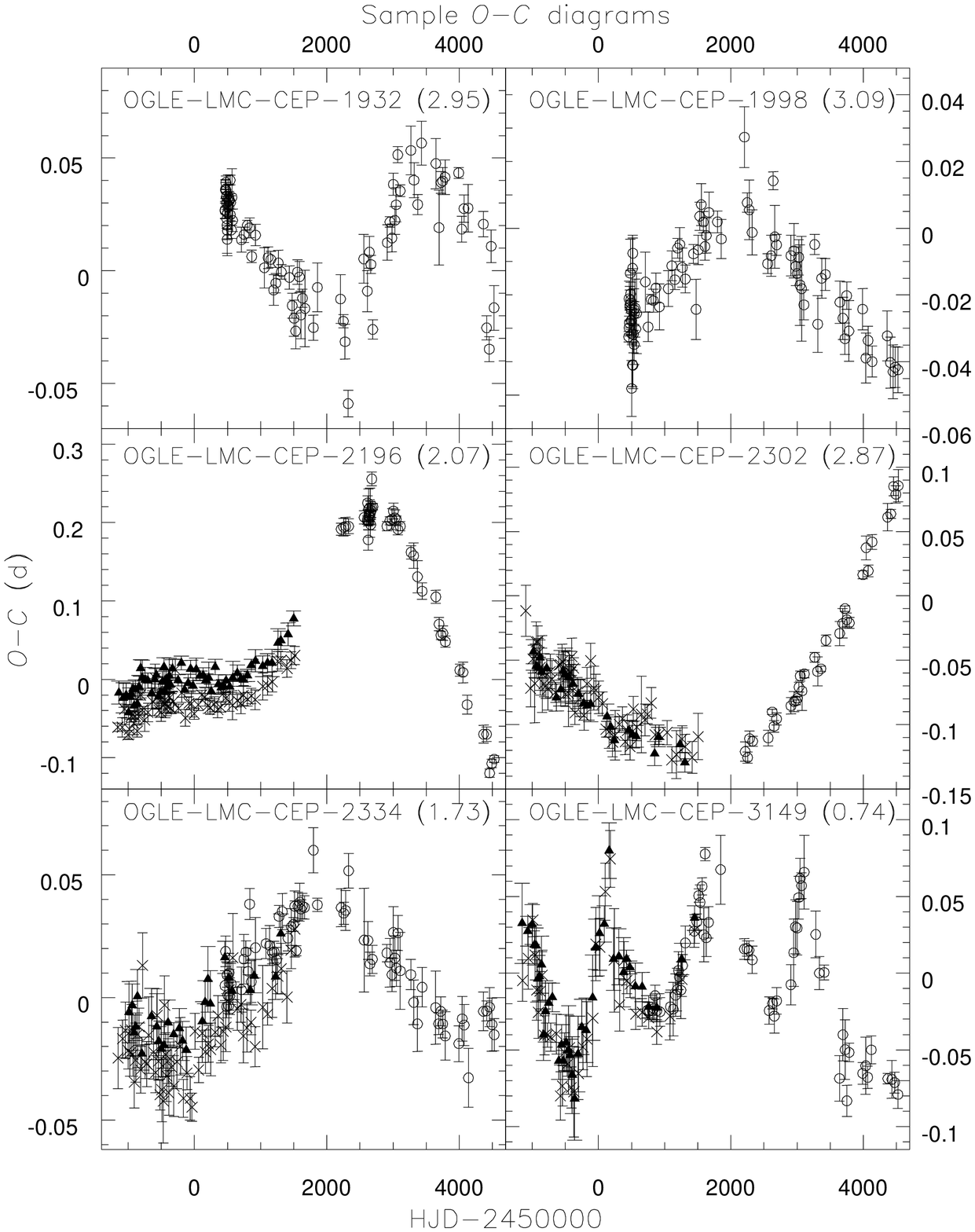}
\FigCap{Same as Fig.~6. OGLE-LMC-CEP-1932 shows nonlinear period 
changes even though only OGLE data are used. OGLE-LMC-CEP-2196 is one of
the stars which are subject to epoch counting error. $O-C$ diagram of
OGLE-LMC-CEP-3149 shows changes with quasi-period $\approx1570$~d long.}
\end{figure}
Figs.~6 and 7 show examples of the $O-C$ diagrams for Cepheids with
determined period changes. Visual inspection of all constructed $O-C$
diagrams revealed that there may be only a few out of 1515 Cepheids
observed by MACHO and OGLE that show parabolic diagrams without any
additional variability. OGLE-LMC-CEP-1119 (LMC\_SC11-338308) was one of the
first overtone Cepheids analyzed by Moskalik and Ko³aczkowski (2008). The
impression one may have after examination of their phase variation graphs
is that the phases change in parabolic or sinusoidal way. The $O-C$ diagram
for OGLE-LMC-CEP-1119 presented in Fig.~6 clearly shows that this statement
is not always true. We also found period changes for four 1O Cepheids for
which they found amplitude changes only. We confirmed period changes for
other first overtone Cepheids listed by Moskalik and Ko³aczkowski (2008,
Table~4).

\Section{Summary and Conclusions}
We have used two different methods to find period changes of single mode
Cepheids. Both gave similar results. Numerical test proved that Pilecki \etal
(2007) error estimation method works well for sinusoidal light curves and
is sufficient for other objects.

Positive and negative period changes show the same dependence on period
(Figs.~3 and 4) with very short timescales of these changes. The sample of
Cepheids changing period shows the same distribution of periods and $(V-I)$
colors as all Cepheids pulsating in the same mode. 1O Cepheids were found
changing period 2.7 times more often than the F pulsators. The timescales
of the found period changes group closer to thermal timescale for 1O than
for F pulsators. Since we have found only some random effects, not the
evolutionary ones, we suspect the F pulsation periods to be more stable than
the 1O ones. Period change rates have opposite signs more often than the same
signs in the MACHO and OGLE data which partly overlap. It may be a result
of some random changes which typical duration is a few thousand days.
Decades long $O-C$ diagrams show similar changes (Berdnikov
1994, Berdnikov and Pastukhova 1994, Berdnikov \etal 1997, Berdnikov \etal
2004) and we conclude that they dominate $O-C$ diagrams for short time base
-- up to 30--40 years. This effect also affects periods derived from a few
thousand day long survey and hence period change rates found by
Pietrukowicz (2001, 2002) do not reflect the evolutionary changes only. Our
results show how difficult is exact period change rate determination. The
way in which the found period changes are connected to Blazhko effect is subject for
further analysis.

Another problem arises for double or triple mode pulsators. The period
change rates can have opposite signs or different and changing rates
(Moskalik and Ko³acz\-kowski 2008), thus simple application of Fourier
fitting with constant and equal in each mode timescale of these changes may
cause spurious results as probably was the case for Moskalik and
Dziembowski (2005) triple mode Cepheids for which seismic models suggested
these objects are in the Hertzsprung gap.

\Acknow{Author is grateful to Prof.\ W.\ Dziembowski, I.\ Soszyñski, 
D.\ Fabrycky and B. Pilecki for fruitful discussions. This work was partly
supported by MNiSW grant NN203293533 to I.\ Soszyñski.

This paper utilizes public domain data obtained by the MACHO Project,
jointly funded by the US Department of Energy through the University of
California, Lawrence Livermore National Laboratory under contract
No.~W-7405-Eng-48, by the National Science Foundation through the Center
for Particle Astrophysics of the University of California under cooperative
agreement AST-8809616, and by the Mount Stromlo and Siding Spring
Observatory, part of the Australian National University.}

\end{document}